\documentclass[amssymb,amsmath,aps,prl,twocolumn,superscriptaddress,showpacs]{revtex4}

\usepackage{amssymb}
\usepackage{epsfig}
\usepackage{graphicx}

\begin{document}
\title{Evidence for Dynamic Phase Separation and High Energy Spin Dynamics in underdoped La$_{2-x}$Sr$_x$CuO$_4$ from Resonant Inelastic X-Ray Scattering}

\author{L. Braicovich}
\affiliation{CNR-INFM Coherentia and Soft, Dipartimento di Fisica,
Politecnico di Milano, I-20133 Milano, Italy}

\author{J. van den Brink}
\affiliation{Leibniz-Institute for Solid State and Materials
Research Dresden, D-01171 Dresden, Germany}
\affiliation{Institute-Lorentz for Theoretical Physics, Universiteit
Leiden, NL-2300 RA Leiden, The Netherlands}
\affiliation{Institute for Molecules \& Materials, Radboud
Universiteit Nijmegen, NL-6500 GL Nijmegen, The Netherlands}

\author{V. Bisogni}
\affiliation{European Synchrotron Radiation Facility, Bo\^{\i}te
Postale 220, F-38043 Grenoble, France}

\author{M. Moretti Sala}
\affiliation{CNR-INFM Coherentia and Soft, Dipartimento di Fisica,
Politecnico di Milano, I-20133 Milano, Italy}

\author{L.J.P. Ament}
\affiliation{Leibniz-Institute for Solid State and Materials
Research Dresden, D-01171 Dresden, Germany}
\affiliation{Institute-Lorentz for Theoretical Physics, Universiteit
Leiden, NL-2300 RA Leiden, The Netherlands}

\author{N.B. Brookes}
\affiliation{European Synchrotron Radiation Facility, Bo\^{\i}te
Postale 220, F-38043 Grenoble, France}

\author{G.M. De Luca}
\affiliation{CNR-INFM Coherentia, Dipartimento Scienze Fisiche,
Universit\`{a} di Napoli ''Federico II'', I-80126 Napoli, Italy}

\author{M. Salluzzo}
\affiliation{CNR-INFM Coherentia, Dipartimento Scienze Fisiche,
Universit\`{a} di Napoli ''Federico II'', I-80126 Napoli, Italy}

\author{T. Schmitt}
\affiliation{Swiss Light Source, Paul Scherrer Institut, CH-5232
Villigen PSI, Switzerland}

\author{G. Ghiringhelli}
\affiliation{CNR-INFM Coherentia and Soft, Dipartimento di Fisica,
Politecnico di Milano, I-20133 Milano, Italy}

\date{\today}

\begin{abstract}
We probe the collective magnetic modes of La$_2$CuO$_4$ (LCO) and underdoped La$_{2-x}$Sr$_x$CuO$_4$ (LSCO) by momentum resolved Resonant Inelastic X-ray Scattering (RIXS) at the Cu $L_3$ edge. In LCO the single
magnon dispersion measured by RIXS coincides with the one determined by inelastic neutron scattering. For LSCO the spin dynamics shows a branch dispersing up to $\sim$400~meV coexisting with a  branch at lower energy. Only the latter has been observed with neutrons so far and is considered a key signature of doped cuprates. The presence of the high-energy branch indicates that LSCO is in a dynamic inhomogeneous spin state.
\end{abstract}


\maketitle

{\it Introduction.}
Since the early days of high $T_{\rm c}$ superconductivity, it has been surmised from a theoretical viewpoint that electronic inhomogeneity plays an essential role in the cuprates, possibly in the form of stripes of holes and spin~\cite{Kivelson03}. Later phase separation was observed experimentally~\cite{Kivelson03,Tranquada07},  revealing the coexistence of patches of distinct electronic phases not only in the form of stripes, but also as charge segregated islands. The central question is whether such spatial inhomegeneity is merely a consequence of static disorder induced by, for instance, the dopants, or if these materials are intrinsically and dynamically phase separated, and extrinsic static potentials just act as pinning centers for the fluctuating charge segregated islands or stripes. We address this question using momentum resolved Resonant Inelastic X-Ray Scattering (RIXS) to probe the collective modes of underdoped La$_{2-x}$Sr$_x$CuO$_4$ (LSCO).

\begin{figure}
\includegraphics[angle=0,width=1.0\columnwidth]{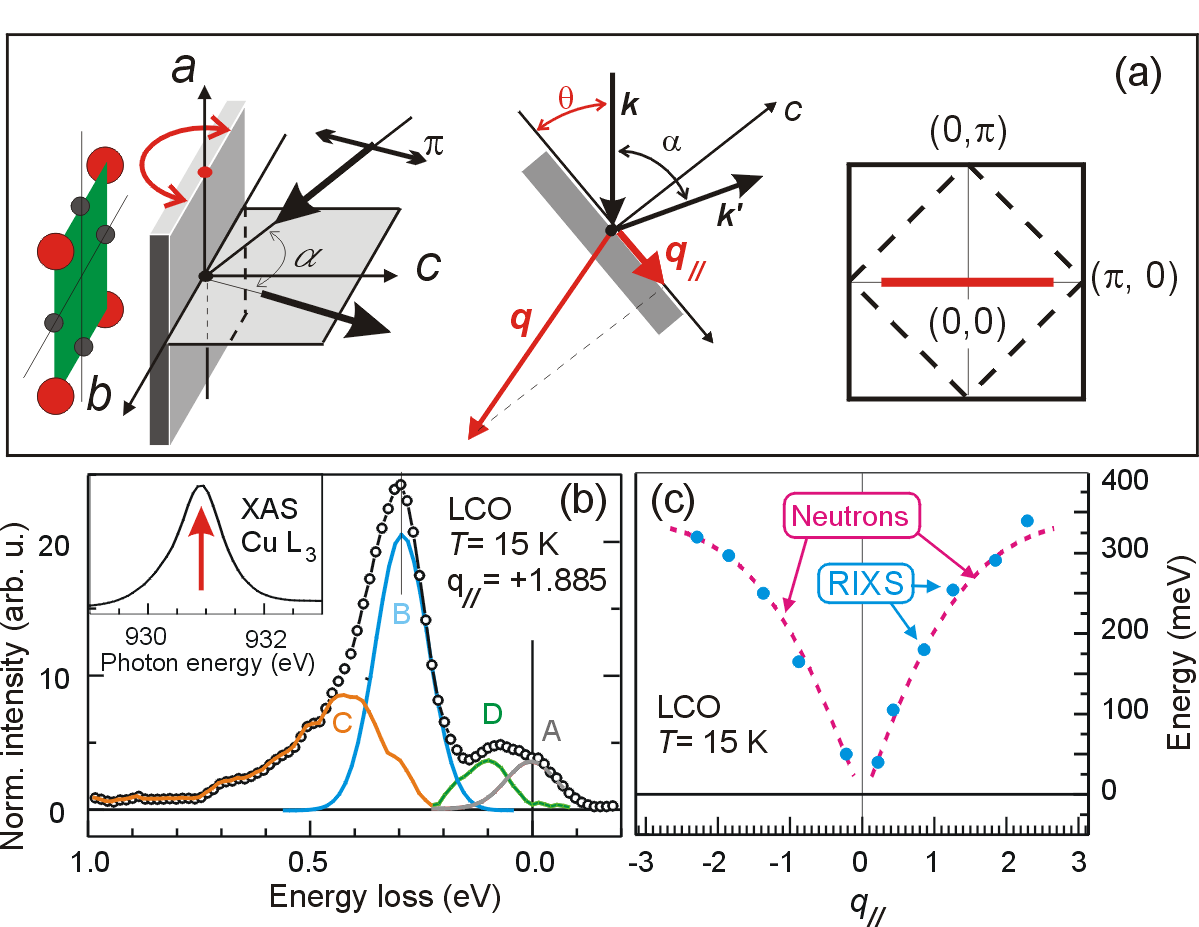}
\caption{(Color on-line) Panel a, lay-out of the experimental set-up and indication of the reciprocal space region (red line) spanned inside the first Brillouin zone. Panel b, decomposition of the LCO spectrum at $q_\parallel$ = 1.85: elastic (A, in grey) and single magnon peaks (B, blue); multiple magnon (C, orange) and optical phonons (D, green) spectral features. In the inset the $L_3$ absorption spectrum, red arrow indicates excitation energy. Panel c, single magnon dispersion determined by RIXS (this work, blue dots) and by inelastic neutron scattering (Ref.~\onlinecite{Coldea01}, dashed purple line).}
\label{fig1}
\end{figure}

RIXS is a unique spectroscopic technique that provides both momentum and energy resolved information on spin and charge dynamics and is, at the same time, element specific~\cite{Kotani01}. We use RIXS at the copper $L_3$ edge, where monochromatic incident radiation ($\bf k$, $h\nu_{in}$) excites a $2p$ core electron resonantly into a Cu $3d$ empty state. This intermediate state quickly decays again, typically within 1-2 femtoseconds~\cite{Keski74}. We measure the energy $h\nu_{out}$ and the momentum ${\bf k}^\prime$ of the photons that are emitted. Thus we determine both the energy $E = h\nu_{in}$-h$\nu_{out}$ and the momentum ${\bf q} ={\bf k}-{\bf k}^\prime$ of the excitation left behind in the sample. The schematics of the experiment is shown in Fig.~\ref{fig1}a; we control the transferred momentum $q_\parallel$ in the $ab$ basal plane  by rotating the sample (angle $\theta$).

In the RIXS process a $3d$ valence electron can, in principle, flip its spin, using the angular momentum carried by the scattered photon. Whether this process is allowed depends on the symmetry of the material under consideration and on the absorption edge. For $K$ edge RIXS only double-spin flip scattering is allowed and on cuprates such as LCO and Nd$_2$CuO$_4$ it was indeed observed by Hill and coworkers~\cite{Hill08}.  In $L_3$ edge RIXS on NiO direct spin-flip scattering is allowed and was recently observed~\cite{Ghiringhelli09}, whereas for cuprates the situation is very peculiar: direct spin-flip transitions are possible only if the copper spin has a non-zero projection in the $xy$ plane~\cite{Ament09}. This corresponds to the actual spin structure of layered cuprates, for which Cu $L_3$ RIXS thus gives access to the fundamental spin excitations.

In undoped antiferromagnetic LCO we use Cu $L_3$ RIXS to measure the magnon dispersion, which we find to coincide with the one obtained with inelastic neutron scattering. Subsequently we investigate underdoped superconducting LSCO and uncover a high-energy branch in the excitation spectrum that coexists with a less-dispersive branch at lower energy. This signals that LSCO is in a dynamically inhomogeneous state, possibly a stripe liquid with coherent spin dynamics up to 400~meV.

{\it Experimental method.}
The LCO and the LSCO samples were 100 nm thick films grown by pulsed laser deposition on the (001) surface of SrTiO$_3$. For LSCO the effective hole density was obtained from the transport measurements that indicated $T_\textrm{c}$=21.5~K. The RIXS experiment was performed at the ADRESS beam line of the Swiss Light Source (Paul Scherrer Institute, Switzerland) using the high resolution SAXES spectrometer \cite{Ghiringhelli06}. The incident x rays were tuned at $\sim$931~eV corresponding to the maximum of the Cu $L_3$ absorption peak. The beam line plane grating monochromator and the spherical grating spectrometer contributed approximately equally to the combined energy resolution $\Delta$E $\leq$ 140 meV.  In Fig.~\ref{fig1}a the experimental lay-out is shown. The crystallographic axes of the sample are ($a$,$b$,$c$) and $\pi$ indicates that the incident x rays are linearly polarized parallel to the scattering plane. $q_\parallel$ is the component parallel to the $ab$ plane of the transferred momentum \textbf{q}. As the photon momentum {\bf k} is dictated by the energy of the Cu $2p\rightarrow3d$ resonant transition, the scattering angle $\alpha$ determines the maximum reachable $q_{\parallel}$. The inset shows the 2D Brillouin zone with the red line indicating the region explored in the experiment.  Negative transferred momentum $q_\parallel$ corresponds by convention to small values of $\theta$, i.e.~incidence near grazing. Each spectrum is the result of 30 minutes total accumulation given by the sum of 6 spectra of 5 minutes. The exact position of the zero on the energy loss scale was determined by measuring for each $q_\parallel$ a non resonant spectrum from polycrystalline graphite.  The spectra were measured with $\alpha = 90^\circ$ (130$^\circ$) for the smaller (larger) $|q_\parallel|$.

The 140 meV energy resolution of our instrumentation allows to use this newly discovered RIXS channel to map out the magnon dispersion in undoped cuprates. Details will be discussed elsewhere~\cite{Ghiringhelli}: here we will use these novel results only to show that in undoped cuprates the magnon dispersion measured using $L_3$ RIXS coincides with the one obtained with inelastic neutron scattering. A typical measured RIXS spectrum of LCO is presented in Fig.~\ref{fig1}b: the raw spectrum is decomposed into an elastic peak A, a single magnon peak B, a high energy feature C and a low energy ($\sim 90$ meV) peak D. The latter is assigned to a well-known optical phonon~\cite{Padilla05}, and feature C is due to higher-order magnetic excitations, namely bimagnons.  In Ref.~\onlinecite{Braicovich09} the lower experimental resolution was insufficient to resolve features C and D for LCO and CaCuO$_2$, and the dispersing features had been assigned mainly to bimagnons. The assignment of the dominant peak B to a single magnon excitation is bolstered by the empirical fact that its dispersion from (0,0) to ($\pi$,0) is in perfect agreement with neutron scattering data~\cite{Coldea01}, as is shown in Fig.~\ref{fig1}c.

\begin{figure}
\includegraphics[angle=0,width=1.0\columnwidth]{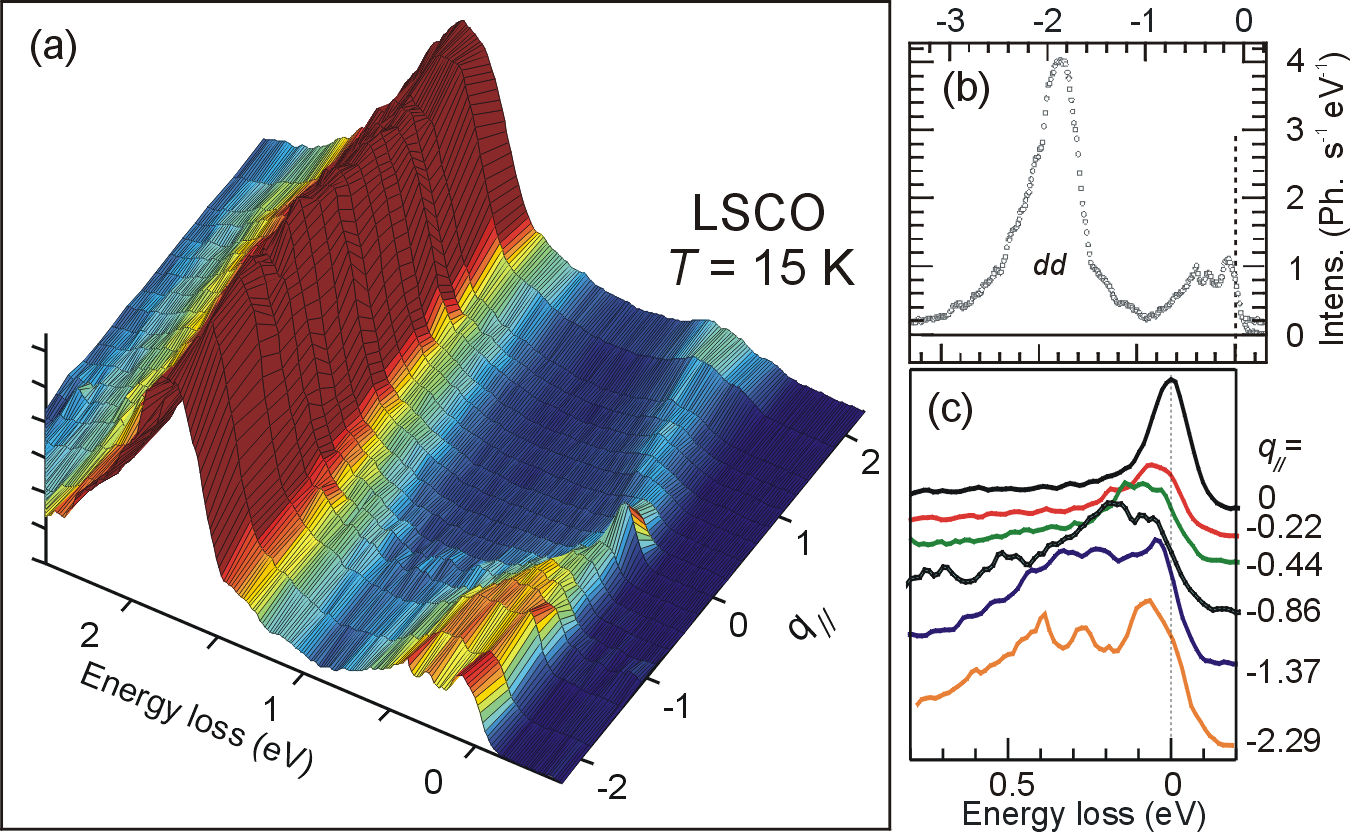}
\caption{(Color on-line) Overview of the RIXS results for LSCO at $T = 15$ K. Panel a, 3D representation of the whole dataset, where the intensity (normalized to the integral of the $dd$ excitations) is plotted as function of the energy loss and of the transferred momentum  $q_\parallel$. Panel b, example of one spectrum including the $dd$ excitations ($q_\parallel$=-2.29), Panel c, selected spectra in the low energy spectral region dominated by magnetic excitations.
}\label{fig2}
\end{figure}

{\it RIXS response of LSCO.} Having demonstrated the power of RIXS to directly probe spin dynamics in the undoped cuprates, we now apply it to the underdoped high $T_\textrm{c}$ superconductor LSCO with a hole concentration $p = 0.08$. The momentum resolved spectra in Fig.~\ref{fig2} cover a large part of the Brillouin zone along the (-$\pi$,0) $\rightarrow$ ($\pi$,0) direction, namely the $q_\parallel$=[-2.29,2.29] range. The peaks at high energy loss are non-dispersive $dd$ excitations, which are very similar to those of LCO. At lower energy, on the contrary, the magnon region of the spectrum is markedly changed by doping.

\begin{figure}
\includegraphics[angle=0,width=0.9\columnwidth]{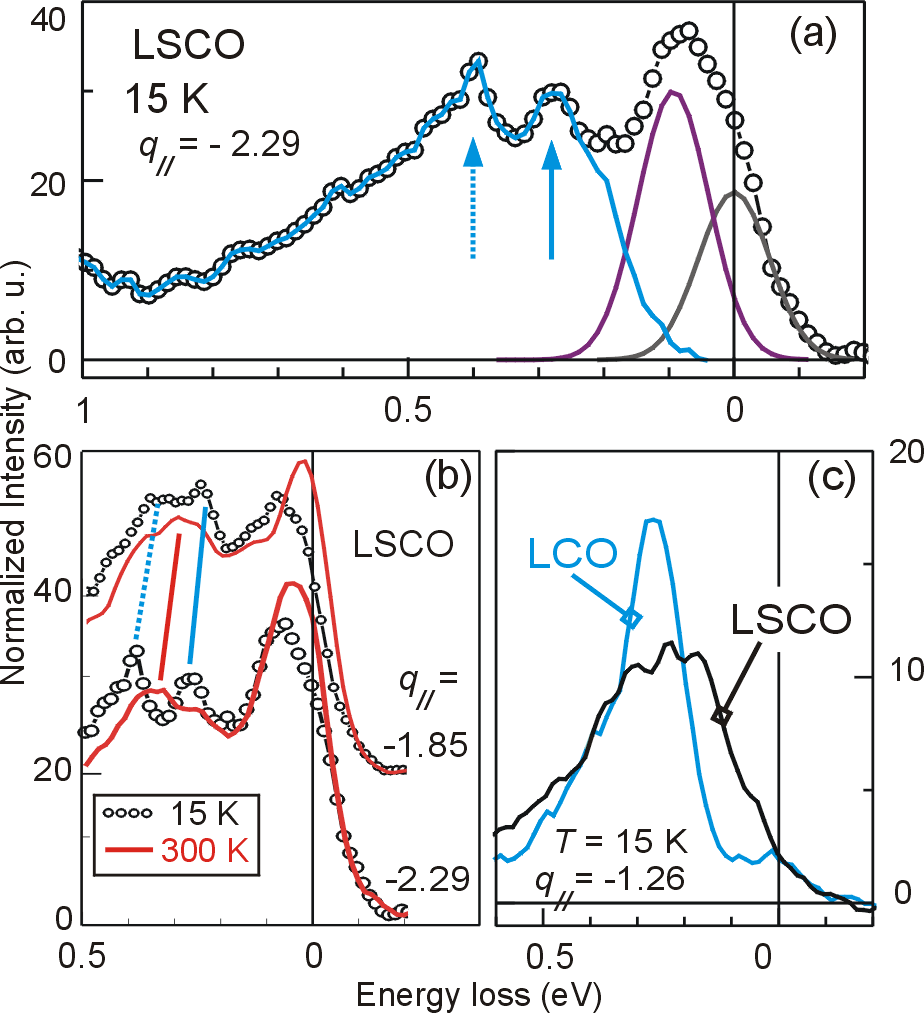}
\caption{(Color on-line) Panel a, example of RIXS spectrum decomposition for underdoped LSCO in the low energy region: elastic peak (grey), phonon peak (purple) and magnetic excitations (blue). Panel b, temperature effects for $q_\parallel$=-2.29 and -1.85: the single peak structure at room $T$  splits into two peaks at low $T$. Panel c, comparison between LCO and LSCO for $q_\parallel$=-1.26 at low temperature. }\label{fig3}
\end{figure}

\begin{figure}
\includegraphics[angle=0,width=1.0\columnwidth]{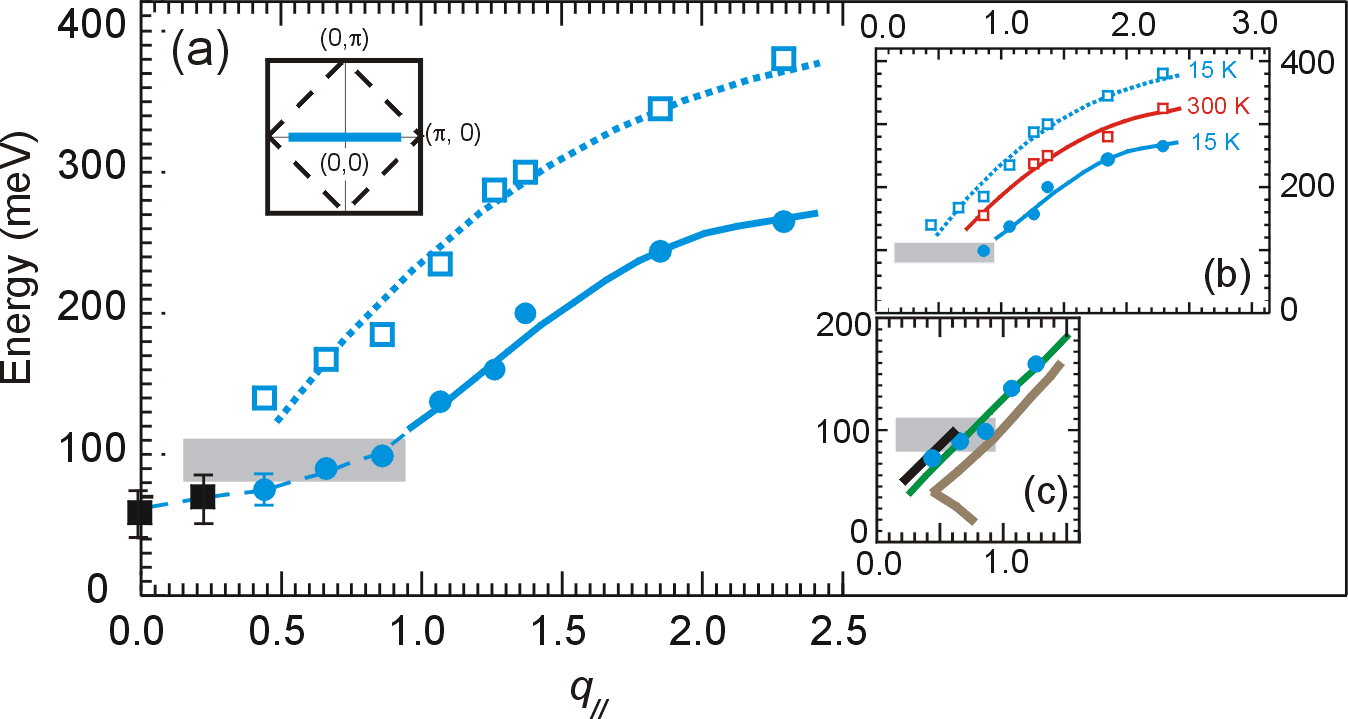}
\caption{(Color on-line) Panel a, dispersion of the two magnetic excitations in underdoped LSCO at $T = 15$ K: one higher energy LCO-like branch (open blue squares) and one lower energy branch (full blue circles). At small $q_\parallel$ (black squares) the separation into subcomponents, if any, is not resolvable. In the shaded area phonon scattering influences the data analysis. The dotted blue line is the LCO dispersion taken from Fig 1(c) and multiplied by 1.09. The solid blue line is a guide to the eye. Panel b, dispersion at room $T$ (red symbols) and low $T$ (blue symbols). Panel c, low energy branch measured with RIXS near (0,0) (blue circles) compared to the neutron data close to ($\pi$,$\pi$) (black line for hole density $p=0.05$, green line for $p$=0.085, brown line for $p= 0.16$).
}\label{fig4}
\end{figure}

Already the three-dimensional intensity plot of Fig.~\ref{fig2}a clearly shows that, when approaching the Brillouin zone boundary, the magnetic excitation spectrum contains multiple \emph{dispersing} peaks. One should note that, after normalization to the $dd$ peak, the spectra are less intense for positive than for negative values of $q_\parallel$. This asymmetry, also clearly present in LCO, is an additional and independent indication for the magnetic character of the dispersing modes. In fact in LCO the asymmetrical intensity fully agrees with the theoretical cross section of Ref.~\onlinecite{Ament09} for pure spin excitations. Thus the permanence of the asymmetry in LSCO is a very strong evidence that also LSCO spectra are dominated by spin dynamics. Although unavoidably contributing to the LSCO spectra to some extent, charge (electron-hole) excitations play a secondary role in the underdoped case.  To fully exploit the RIXS information on spin dynamics we made a detailed decomposition of each spectrum. Our data analysis is illustrated in Fig.~\ref{fig3}a for $q_\parallel$=-2.29. After subtracting, in analogy to LCO, the elastic line and phonon losses, the resulting magnetic part of the spectrum is characterized by two separate peaks. The double peak structure of the magnetic spectrum is observed at low temperature (15 K in the present experiment) while the peaks amalgamate at room temperature~\cite{Machtoub05},  (see Fig~\ref{fig3}b). For $|q_\parallel|$$\ge$0.44 the two contributions to the magnetic scattering intensity are discernable. To highlight the doping induced changes in the spin dynamics, in Fig.~\ref{fig3}c the RIXS spectra of LSCO and LCO are compared at $q_\parallel$=-1.26. It is apparent that upon hole doping the magnon peak of LCO looses intensity and a lower energy feature appears whilst also the higher energy part changes. The resulting dispersion of the two magnetic branches is shown in Fig.~\ref{fig4}a. The upper branch, reaching 400 meV, is very close to the magnon dispersion of the parent compound LCO. A detailed analysis of the spectra shows that within $\sim$25~meV, the two branches are symmetric in $q_\parallel$, whereas the assignment of two separate branches becomes ambiguous for $|q_\parallel|$$<$0.44. At room temperature the two peaks do not separate and the dispersion of the resulting magnetic feature lies in between the two low temperature branches, as shown in Fig.~\ref{fig4}b.

The lower energy magnetic branch also appears in recent neutron data by Lipscombe {\it et al}.~\cite{Lipscombe09}  on underdoped LSCO with the same hole concentration ($p$=0.085). The branch disperses up to 200 meV, see Fig.~\ref{fig4}c, which indeed corresponds to our lower magnetic RIXS branch. The dispersion of the magnetic RIXS branch is also precisely in between the one determined with neutron scattering  on LSCO for $p = 0.05$ \cite{Goka03} and the one for $p$=0.16 by Vignolle {\it et al}.~\cite{Vignolle07}. At smaller $q_\parallel$ we observe a magnetic RIXS signal within the limits of our resolving power near the neck of the famous ``hour glass'' dispersion seen by neutron scattering~\cite{Tranquada07,Vignolle07,Tranquada04,Christensen04}. One must keep in mind, though, that in RIXS we are measuring momenta away from (0,0) and the neutron data are taken around ($\pi$,$\pi$). These are equivalent points in the Brillouin zone only in the presence of long-range antiferromagnetic order. Irrespectively of this, the characteristics of the lower branch of the magnetic excitation spectrum that we observe in RIXS nicely agree with the inelastic neutron scattering branch observed in the same system around ($\pi$,$\pi$). In contrast we have discovered a higher-energy magnetic branch dispersing up to $\sim$400 meV in a doped cuprate superconductor. As indicated above, this branch is very close to the magnon dispersion of the LCO showing that high energy spin dynamics is present also in underdoped LSCO, a fact not observed with neutrons yet. Observing in underdoped LSCO both a low- and high-energy magnetic branch suggests the presence of an inhomogeneous magnetic state: although the energy is similar, the spectral shape of the higher energy feature in LSCO is different than in LCO, indicating that the phase separation takes place at a microscopic scale.

To sustain a coherent LCO-like magnetic excitation propagating as shown by dispersion with $q_\parallel$ around 0.44, corresponding to a wavelength of $\sim$7 lattice parameters (i.e. $\sim$27 \AA), a magnetic patch must have a linear extend that is considerably longer. This extend can be associated with the dimensions of the patches (either stripes or islands) on the 1-2 fs time scale of $L$ edge RIXS. This is consistent with the notion that RIXS is a very fast probe and can, as it were, take an ``instantaneous'' snapshot of slow fluctuating magnetic states and associated excitations. Less rapid methods are thus expected to observe smaller island sizes. Indeed on the basis of nuclear quadrupole resonance measurements Singer {\it et al}.~\cite{Singer02}  report magnetic patch size of 30 \AA \ in LSCO. Also with scanning tunneling microscopy, another very slow probe, Kohsaka {\it et al}.~\cite{Kohsaka04} observe in the related cuprate superconductor Ca$_{2-x}$Na$_x$CuO$_2$Cl$_2$  ($x$ around 0.08) separate metallic and insulating areas of approximately 20~\AA~ in diameter.

The present experimental data are compatible with a model of stripe patches fluctuating along the crystallographic axis of a twinned sample (i.e.~a kind of stripe liquid). According to the $t$-$J$ model calculations by Seibold and Lorenzana~\cite{Seibold06} for static stripes, there are two magnetic branches separated by about 100 meV, of which the higher branch is a remnant of LCO as it corresponds to magnetic excitations parallel to the stripe direction. The presence of these two excitations is consistent with our data. However, we do not observe static stripe formation, which suggests that at low temperature underdoped LSCO is a stripe liquid in which the directions perpendicular and parallel to the stripes are macroscopically indistinguishable. An additional indication for the dynamical nature of this inhomogeneous magnetic state is that high energy van Hove singularities at $q_\parallel$=0, which are typical of static stripes~\cite{Seibold06}, are absent; instead we observe a rather flat distribution of spectral intensity up to about 0.9 eV (see in Fig.~\ref{fig2}c the spectra at $q_\parallel$=0 and -0.22).

{\it Conclusions.} The present finding of a high-energy branch in the excitation spectrum of an underdoped high $T_\textrm{c}$ superconductor coexisting with a less-dispersive branch at lower energy signals that LSCO is in a dynamically inhomogeneous state, possibly a stripe liquid with spin dynamics up to 400~meV. In more general terms, our experimental findings demonstrate the power of newest generation high-resolution momentum dependent RIXS to probe the spin dynamics of high $T_{\rm c}$ superconductors. This paves the way to further clarify the interplay of the observed magnetically inhomogeneous state with superconductivity. To do so, further cross-fertilization between RIXS and neutron spectroscopy will be of paramount importance.

\begin{acknowledgements}
This work was performed at the ADRESS beam line of the SLS (PSI) using the SAXES spectrometer developed jointly by Politecnico di Milano, SLS and EPFL. The authors acknowledge Tom Devereaux, Marco Grioni, Steve Kivelson, George Sawatzky, Z.-X. Shen, and Jan Zaanen for important exchanges of ideas; and Milan Radovic and Fabio Miletto Granozio for providing the LCO sample.
\end{acknowledgements}

\end{document}